\documentstyle[multicol,aps,prb]{revtex}

\begin{document}
\draft
\title{ Disappearance of Ensemble-Averaged Josephson Current \\
in Dirty SNS Junctions 
of \bbox{\boldmath d}-wave Superconductors}
\author{Yasuhiro Asano$^\ast$}
\address{
Department of Applied Physics, Hokkaido University, Sapporo 060-8628, Japan.\\}
\date{\today}
\maketitle
\begin{abstract}
We discuss the Josephson current
in superconductor / dirty normal conductor /
superconductor junctions, where the superconductors 
have $d_{x^2-y^2}$ pairing symmetry. 
The low-temperature behavior of the Josephson current
depends on the orientation angle between the crystalline axis and 
the normal of the junction interface.
We show that the ensemble-averaged Josephson current vanishes when 
the orientation angle is $\pi/4$ and the normal conductor is in the
diffusive transport regime.   
The $d_{x^2-y^2}$-wave pairing symmetry is responsible for this fact.
\end{abstract}

\hspace{0.5cm}

\begin{multicols}{2}
\narrowtext

\section{introduction}

In recent years, the Josephson effect between the anisotropic 
superconductors
has attracted much attention because the 
high-$T_c$ superconductors might have the $d_{x^2-y^2}$ pairing 
symmetry~\cite{sigrist,wollman}.
In anisotropic superconductors, the sign of the
pair potential depends on the direction of a quasiparticle's motion.
As a consequence, the zero-energy states (ZES)~\cite{hu} are formed
at the normal metal / superconductor (NS) interface when the 
potential barrier at the interface is large enough.  
The ZES are sensitive to the orientation angle between the
crystalline axis of the high-$T_c$ superconductors 
and the normal of the junction interface. 
The zero-bias conductance-peak which is due to the ZES 
is observed in conductance spectra of 
N / I / $d$-wave superconductor junctions and
the peak-height is maximum  when the 
orientation angle is $\pi/4$~\cite{iguchi}.
Here I is the insulator.
In SIS junctions, the ZES dominate the dc Josephson effect 
and the low-temperature anomaly in 
the Josephson current has been discussed  
in a number of theoretical 
works~\cite{barash,tanaka,samanta,riedel,golubov,fogelstrom,zhang,zhu}.
In experiments, it seems to be difficult to fabricate clean SIS and SNS
junctions. Thus it is important to understand the effects of disorder 
on the Josephson current. So far it has been pointed out
that the roughness at the interface suppresses the low-temperature anomaly
of the Josephson current in SIS junctions~\cite{barash,riedel,golubov}.

In this paper, we study the dc Josephson effect in 
superconductor / normal conductor / superconductor junctions, 
where the superconductors have the $d_{x^2-y^2}$-wave pairing symmetry
and the normal metal is in the diffusive transport regime.
We analytically derive the 
Josephson current in dirty SNS junctions based on
a formula~\cite{furusaki} in which 
the Josephson current is calculated from the two Andreev 
reflection coefficients~\cite{andreev}.
We show that the ensemble-averaged Josephson current vanishes
when the orientation angle is $\pi/4$. 
The analytic results are confirmed by a numerical
simulation by using the recursive Green function method~\cite{asano}.
Throughout this paper we take units of $k_B=\hbar=1$,
where $k_B$ is the Boltzmann constant.

This paper is organized as follows. In Sec.~2, we derive the 
general expression of the Josephson current in SNS junctions. 
In Sec.~3, we discuss the Josephson current in dirty SNS junctions. 
The discussion is given in Sec.~4. We summarize this paper 
in Sec.~5.  

\section{Josephson Current Formula}

Let us consider the two-dimensional SNS junction as shown in 
Fig.~\ref{system}, where 
the length of the normal metal is $L_N$ and the width of the junction is $W$,
respectively. The pair potential in the momentum space is schematically
depicted in the superconductors. 
In the $d_{x^2-y^2}$-wave superconductor, the $a$-axis points the 
direction in which the amplitude of the pair potential takes its maximum.
The angle between the $x$ direction and the $a$-axis is 
$\alpha_L$ ($\alpha_R$) in the superconductor on the left (right) hand side. 
We assume that the effective mass ($m$) 
of an electron and the Fermi energy ($\mu_F= k_F^2/2m$) 
are common in the superconductors and the normal conductor, 
where $k_F$ is the Fermi wavenumber. 
We describe the SNS junction by using the Bogoliubov-de Gennes (BdG) 
equation~\cite{degennes},
\begin{eqnarray}
&\int& d{\bf r}'
 \left[
  \begin{array}{cc} 
  \delta({\bf r}-{\bf r}') h_0({\bf r}') & \Delta({\bf r},{\bf r}') \\
  \Delta^\ast({\bf r},{\bf r}') & -\delta({\bf r}-{\bf r}') h_0({\bf r}')
  \end{array}
 \right]
 \left[
 \begin{array}{c}
  u({ \bf r}')\\
  v({ \bf r}')
 \end{array}
 \right] \nonumber \\
& &= E
 \left[
 \begin{array}{c}
  u({ \bf r})\\
  v({ \bf r})
 \end{array}
 \right],
\end{eqnarray}
\begin{eqnarray}
h_0({\bf r})&=& -\frac{\nabla^2}{2m} + U({ \bf r}) - \mu_F, \\
U({ \bf r}) &=& V_b\left[ \delta(x)+\delta(x-L)\right] + \sum_{i=1}^{N_{i}}
 v_I \delta( {\bf r}-{\bf r}_i), \label{pot} 
\end{eqnarray}
where the first term in Eq.~(\ref{pot}) is the barrier potential 
at the NS interface and the second term is the impurity potential in the normal
metal, respectively. 
The energy of a quasiparticle ($E$) is measured from the chemical potential
 of the junction.
The pair potential can be described by the Fourier representation,
\begin{equation}
\Delta({\bf R}, {\bf r}-{\bf r}') = \frac{1}{(2\pi)^{2}}
 \int \int d{\bf k} \, \Delta({\bf k})\, 
{\rm e}^{i{\bf k} ({\bf r}-{\bf r}')},
\end{equation}
where we assume that the pair potential is uniform in the superconductors
and neglect the dependence of $\Delta$ on ${\bf R}=({\bf r}+{\bf r}')/2$.
The $s$-wave superconductor is characterized
by $\Delta({\bf k}) = \Delta {\rm e}^{{\rm i}\varphi_j}$.
The $d_{x^2-y^2}$-wave superconductor is characterized by
\begin{eqnarray}
\Delta({\bf k}) &=& \Delta {\rm e}^{{\rm i}\varphi_j} \cos(2\alpha_j - 2\gamma),\\
{\rm e}^{{\rm i}\gamma}&=&\cos\gamma + {\rm i}\sin\gamma 
=(k_x + {\rm i} k_y)/k_F,
\end{eqnarray} 
where $\Delta$ is the uniform amplitude of the pair potential, 
$\varphi_j$ and $\alpha_j$ with ($j=L$ or $R$) are the phase of the pair
potential and the orientation angle, respectively~\cite{tanaka}.
At the Fermi surface, the wavenumber in the $x$ direction is $k_x$
and that in the $y$ direction is $k_y$, respectively. 
In the normal conductor, the pair potential is set to be zero. 

We calculate the Josephson current by using the Furusaki-Tsukada 
formula~\cite{tanaka,furusaki},
\begin{eqnarray}
J&=& e T \sum_{\omega_n} \sum_{k_y}\nonumber \\
 &\times&\left[
\frac{|\Delta^+_L| a_1(k_y,\omega_n)}{\Omega^+_L}
-\frac{|\Delta^-_L| a_2(k_y,\omega_n)}{\Omega^-_L} \right],
\label{jdef} \\
\Delta^{\pm}_j&=&\Delta \cos 2(\alpha_j \mp \gamma),\\
\Omega^{\pm}_j &=& \sqrt{\omega_n^2+ (\Delta^{\pm}_j)^2},
\end{eqnarray}
with $j=L$ or $R$.
Here $T$ and $\omega_n= (2n+1)\pi T$ are the temperature and the Matsubara 
frequency, respectively.
The wavenumber in the $y$ direction $k_y$ specifies the propagating channel 
and the number of propagation channels for each spin is $N_c\equiv Wk_F/\pi$.
The coefficients $a_1(k_y, \omega_n)$ and $a_2(k_y, \omega_n)$ 
are the analytic continuation 
($E\rightarrow i\omega_n$) of the reflection coefficients
of a quasiparticle from the left superconductor to the left superconductor. 
The Andreev reflection coefficient from
the electron (hole) channels to the
hole (electron) channels is denoted by $a_1$ ($a_2$).
In the presence of the time reversal symmetry, the Josephson
current can be decomposed into the series of 
\begin{equation}
J = \sum_{m=1}^\infty J_m \sin( m \varphi ), \label{deco} 
\end{equation}
where $\varphi =\varphi_L-\varphi_R$.
In this paper, we neglect the series 
of $J_m$ for $m \geq 2$. This approximation is justified
in the presence of the high potential barrier
at the NS interface. 
In Fig.~\ref{process} (a), we show the four reflection processes 
of $ a_1$ and $ a_2$ 
which contribute to $ J_1$. 
In order to estimate $ a_1$ and $ a_2$, we calculate the transmission
and the reflection coefficients of the single NS interface for 
fixed $k_y$. 
The sixteen coefficients are obtained from the continuity condition 
of the wavefunction at the NS interface since there are four 
incoming and four outgoing channels for each $k_y$. 
In Appendix A, we show eight transmission and four reflection coefficients 
which are required in the
following calculation.
The Andreev reflection coefficients in Fig.~\ref{process} (a) are estimated
as
\begin{eqnarray}
&a_1^{(1)}&(k_y,\omega_n) =  \sum_{k'_y} t_{SN}^{hh}(k_y,\alpha_L,\varphi_L) 
\; t^h_{k_y,k'_y} 
\; \nonumber \\ 
&\times& r_{NN}^{he}(k'_y,\alpha_R,\varphi_R) \; t^e_{k'_y,k_y}
\;  t_{NS}^{ee}(k_y,\alpha_L,\varphi_L), \label{a11}\\
&a_1^{(2)}&(k_y,\omega_n) =  \sum_{k'_y} t_{SN}^{he}(k_y,\alpha_L,\varphi_L) 
\; t^e_{k_y,k'_y} 
\; \nonumber \\ 
&\times& r_{NN}^{eh}(k'_y,\alpha_R,\varphi_R) \; t^h_{k'_y,k_y}
\;  t_{NS}^{he}(k_y,\alpha_L,\varphi_L), \label{a12}\\
&a_2^{(1)}&(k_y,\omega_n) =  \sum_{k'_y} t_{SN}^{ee}(k_y,\alpha_L,\varphi_L) 
\; t^e_{k_y,k'_y} 
\; \nonumber \\ 
&\times& r_{NN}^{eh}(k'_y,\alpha_R,\varphi_R) \; t^h_{k'_y,k_y}
\;  t_{NS}^{hh}(k_y,\alpha_L,\varphi_L),\label{a21} \\
&a_2^{(2)}&(k_y,\omega_n) =  \sum_{k'_y} t_{SN}^{eh}(k_y,\alpha_L,\varphi_L) 
\; t^h_{k_y,k'_y} \nonumber \\ 
&\times&
\;  r_{NN}^{he}(k'_y,\alpha_R,\varphi_R) \; t^e_{k'_y,k_y}
\;  t_{NS}^{eh}(k_y,\alpha_L,\varphi_L), \label{a22}
\end{eqnarray}
where $t^{e(h)}_{k'_y,k_y}$ is the transmission coefficient
of the electronlike (holelike) quasiparticle in the normal conductor,
and $k'_y$ indicates the propagating channel at the right
NS interface.
The transmission coefficients in the normal metal are described 
by 
\begin{eqnarray}
t^{e}_{k'_y,k_y} &=& {\rm i} v_F\cos\gamma 
 {\rm e}^{-{\rm i}k_F\cos\gamma'L_N}
 \int_{0}^{W} dy \int_{0}^{W}dy' \nonumber \\
&\times&
\, {\cal G}_{\omega_n}(L_N,y'; 0,y) 
 \chi_{k'_y}^\ast(y') \chi_{k_y}(y),\label{te} \\
t^{h}_{k_y,k'_y} &=& -{\rm i} v_F \cos\gamma'
 {\rm e}^{{\rm i} k_F\cos\gamma'L_N}
 \int_{0}^{W} dy \int_{0}^{W}dy'\nonumber\\
&\times&\, {\cal G}_{-\omega_n}(L_N,y'; 0,y) 
 \chi_{k_y}^\ast(y) \chi_{k'_y}(y'),\label{th} 
\end{eqnarray}
where ${\cal G}_{\omega_n}({\bf r},{\bf r}')$ is the Green function
of the normal conductor, $v_F$ is the Fermi velocity, and 
$\chi_{k_y}(y)$ is the
wavefunction in the $y$ direction belonging to the channel specified 
by $k_y$, respectively~\cite{stone}.
In this paper, we assume that the NS interface is sufficiently clean
so that $k_y$ is conserved while the transmission and the 
reflection at the 
interface. 
In $a_1^{(1)}$ in Eq.~(\ref{a11}), 
a quasiparticle wave is initially incident into the normal part
from the left superconductor through the channel specified by $k_y$.
After the Andreev reflection at the right NS interface, 
we assume that the reflected wave  
transmits to the left superconductor through the initial channel 
of $k_y$ because of the retro property of a 
quasiparticle under the time reversal symmetry~\cite{bennaker}.  
The two Andreev reflection coefficients in Eq.(\ref{jdef}) are 
given by $a_1=a_1^{(1)}+ a_1^{(2)}$ and $a_2=a_2^{(1)}+ a_2^{(2)}$, 
respectively.
By using the transmission and the reflection coefficients in Appendix A,
the following equations can be derived,
\begin{eqnarray}
& &\sum_{k_y}\left[
\frac{|\Delta^+_L|}{\Omega^+_L}a_1^{(1)} 
-\frac{|\Delta^-_L|}{\Omega^-_L}a_2^{(2)} \right] 
= 2{\rm i} \sum_{k_y,k'_y} P_1, \\
& &\sum_{k_y}\left[
\frac{|\Delta^+_L|}{\Omega^+_L}a_1^{(2)} 
-\frac{|\Delta^-_L|}{\Omega^-_L}a_2^{(1)} \right] 
= -2{\rm i} \sum_{k_y,k'_y} P_2, 
\end{eqnarray}
with
\begin{eqnarray}
P_1&\equiv&
r_{NN}^{eh}(k_y,\alpha_L,\varphi_L) \:t^h_{k_y,k'_y} \:
r_{NN}^{he}(k'_y,\alpha_R,\varphi_R) \nonumber \\
& &\times \:t^e_{k'_y,k_y}, \\
P_2&\equiv&
r_{NN}^{he}(k_y,\alpha_L,\varphi_L) \:t^e_{k_y,k'_y} \:
r_{NN}^{eh}(k'_y,\alpha_R,\varphi_R) \nonumber \\
& &\times\:t^h_{k'_y,k_y}. 
\end{eqnarray}
The Josephson current in Eq.~(\ref{jdef}) can be written as
\begin{equation}
J= 2{\rm i}
eT \sum_{\omega_n}\sum_{k_y,k'_y}\left( P_1- P_2\right)\label{jgene2}.
\end{equation}
The equation in Eq.~(\ref{jgene2}) corresponds to the reflection processes
shown in Fig.~\ref{process} (b). 
After small algebra, we find the final expression of the Josephson current,
\begin{equation}
J = 4e \sin\varphi T\sum_{\omega_n}\sum_{k_y,k'_y} 
Q(k_y,\alpha_L) Q(k'_y,\alpha_R) \:{\cal T}(k_y,k'_y), \label{jgene}
\end{equation}
with
\begin{eqnarray} 
{\cal T}(k_y,k'_y)&\equiv& t^h_{k_y,k'_y} t^e_{k'_y, k_y}
=t^{e}_{-k_y,-k'_y}t^{h}_{-k'_y,-k_y}, \\
Q(k_y,\alpha_j)&\equiv&
\frac{\cos^2\gamma \Delta^-_jK^+_j}{\Xi_j}, \label{qdef}
\end{eqnarray}
\begin{eqnarray}
\Xi_j&\equiv&
Z^2(\Delta^+_j\Delta^-_j + K^+_jK^-_j) +\cos^2\gamma \Delta^+_j\Delta^-_j,\\
K^\pm_j&\equiv& \Omega^\pm_j-|\omega_n|,
\end{eqnarray}
where 
$Z=mV_b/k_F$ represents
the strength of the barrier potential.
In what follows, we consider the high potential barrier at
the NS interface, (i.e., $ Z\gg 1$).
We note that $Q(k_y,\alpha_j)$ is proportional to the 
Andreev reflection coefficient at the NS interface.
Equations~(\ref{jgene2}) and (\ref{jgene}) are
the general expressions of the Josephson current proportional to
$\sin\varphi$. It is possible to apply these expressions to 
various junctions if the transmission coefficients in the normal
part can be calculated.

\section{Dirty SNS Junctions}
When the normal conductor is in the diffusive transport regime,
$t^{e(h)}_{k'_y,k_y}$ is almost independent of the
propagating channels, 
\begin{equation} 
\langle {\cal T}(k_y,k'_y) \rangle 
\rightarrow \overline{\langle t^e t^h \rangle},\label{iso}
\end{equation}
where $\langle \cdots \rangle$ means the ensemble
average and $\overline {\cdots}$ is the average over the propagation channels.
The transmission coefficients are approximately given by
\begin{eqnarray}
\overline{\langle t^e t^h \rangle} 
&=& \frac{v_F^2}{2N_c^2} \int_0^W dy \int_{0}^{W}  dy'\, 
X(L_N,y; 0,y'),  
\label{aproxtt} \\
X({\bf r}, {\bf r}') &\equiv &
\langle {\cal G}_{\omega_n}({\bf r}, {\bf r}') {\cal G}_{-\omega_n}
({\bf r},{\bf r}') \rangle.
\end{eqnarray} 
In the diffusive regime, the function $X({\bf r},{\bf r}')$ 
for small $\omega_n$ satisfies the diffusion equation,
\begin{equation}
\tau_0\left(2|\omega_n|-D_0 \nabla^2_{\bf r} \right)X({\bf r},{\bf r}')
\approx  2\pi N_0 \tau_0\delta({\bf r}-{\bf r}'), \label{diff}
\end{equation}
where $D_0$, $\tau_0$ and $N_0$ are the diffusion constant, the mean free
time and the density of states at the Fermi energy per unit area for 
each spin degree of freedom, respectively.
The propagating process in the normal conductor is diagrammatically
illustrated in Fig.~\ref{diffusion}, where the cross represents the impurity
scattering. We solve Eq.~(\ref{diff}) and show the results in Appendix B.
The averaged transmission coefficients are
\begin{equation}
\overline{\langle t^e t^h \rangle} = \left(\frac{1}{N_c}\right)^{2} 
g_N \frac{ln}{\sinh ln},
\label{teth}
\end{equation}
where $ln= \sqrt{2n+1}L_N/\xi_D$, $\xi_D= \sqrt{D_0/2\pi T}$ is the 
coherence length, and $G_N\equiv(2e^2/h) g_N$ is the 
conductance in the normal metal, respectively. 
The ensemble averaged Josephson current in dirty SNS junctions is 
calculated by using Eqs.~(\ref{jgene}) and (\ref{teth})
\begin{eqnarray}
\langle J(\alpha_L,\alpha_R) \rangle &=& e
T \sum_{\omega_n} 
N(\alpha_L) N(\alpha_R)
g_N \frac{ln}{\sinh ln}, 
\label{jsns} \\
N(\alpha_j) &\equiv& \int_{-\pi/2}^{\pi/2} d\gamma 
\frac{\cos^3\gamma \Delta^-_jK^+_j}{{ \Xi}_j}, \label{kn}
\end{eqnarray}
where we replace the summation $\sum_{k_y}$ by the integration 
$(N_c/2)\int^{\pi/2}_{-\pi/2}d\gamma\cos\gamma$. 
When the orientation angle are zero, (i.e., $\alpha_{L,R}=0$), 
the ensemble-averaged critical 
current in units of $\pi \Delta_0/eR_J$ 
is given by,
\begin{equation}
\langle{\bar J}(0,0)\rangle = \frac{9}{25} \frac{\Delta}{\Delta_0}
T\sum_{\omega_n} 
\frac{\Delta}{\omega^2_n+\Delta^2}\frac{ln}{\sinh ln}, 
\label{jc1} \\
\end{equation}
where the resistance of the junction is described by 
\begin{equation}
R_J= \left[\frac{2e^2}{h} \frac{4}{9} \frac{g_N}{Z^4}\right]^{-1},
\end{equation}
and $\Delta_0$ is the amplitude of the pair potential at $T=0$.
In the case of the $s$-wave junctions, the numerical factor 9/25 in
Eq.~(\ref{jc1}) is replaced by 1 and 
the expression is identical to that by the Usadel equation~\cite{likharev}.
In the recent studies~\cite{dubos,zaikin} and references
therein, it is pointed out that the results in Ref.~\onlinecite{likharev} 
is not correct in the low temperature regime and the non-sinusoidal
current-phase relation is observed. 
In these studies, the electron transmission at the NS interface is
perfect and the higher harmonics contribute to the Josephson current. 
In our results, the amplitude of the Josephson current 
takes its maximum at $\varphi=\pi/2$ for all temperature regime 
because of the potential barrier at the NS interface.

When $\alpha_{L,R}=\pi/4$, we find that
$N(\pi/4) = 0$, therefore,
\begin{equation}
\langle {\bar J}(\pi/4,\pi/4)\rangle = 0.\label{zero}
\end{equation}
This is
because  
$K^+_{L,R}$ and ${\Xi}_{L,R}$ are even function of $\gamma$, whereas
$\Delta^-_{L,R}=-\Delta\sin 2\gamma$ 
is odd function of $\gamma$ in Eq.~(\ref{kn}). 
The symmetry of the pair potential
is responsible for the disappearance of the ensemble-averaged 
Josephson current.

When the orientation angle is $\pi/8$, 
the averaged critical current results in
\begin{equation}
\langle{\bar J}(\pi/8,\pi/8) \rangle= \frac{9\pi^2}{512} 
\frac{\Delta}{\Delta_0}
 T\sum_{\omega_n} \frac{\Delta}{\omega^2_n+\Delta^2/2}
\frac{ln}{\sinh ln}. 
\label{jc2} 
\end{equation}     
The condition for the
ZES (i.e., $\Delta^+_j\Delta^-_j<0$) is satisfied in the range of 
$\pi/8 \leq |\gamma| \leq 3\pi/8$ in Eq.~(\ref{kn}). 
In the same reason with the case
of $\alpha_{L,R}=\pi/4$, the contribution from
this range becomes zero. 
In Eq.~(\ref{jc2}), the integration in the range of 
$0 \leq |\gamma| \leq \pi/8$ in Eq.~(\ref{kn}) 
contributes to the averaged Josephson current.
We conclude that the ZES do not contribute
to the averaged Josephson current in dirty SNS junctions
of the $d$-wave superconductor.

In Fig.~\ref{result} (a) and (b), we show the numerical 
results of the Josephson current at $\varphi=\pi/2$
by using the recursive Green function method on the two-dimensional 
tight binding model~\cite{asano}, where $\Delta_0=0.01 \mu_F$, 
$L_N=100 a_0$, $W=30 a_0$ and $a_0$
is the lattice constant. The mean free path ($\ell$) is about $6.2 a_0$.
The numerical results for $\alpha_{L,R}=0$ and $\pi/4$ are
shown in Fig.~\ref{result} (a) and (b), respectively.
The dependence of $\Delta$ on $T$ is described by the BCS theory.
The lines are the Josephson current for several samples with different random
configurations and the open circles denote the ensemble average.
In (a), the results for all samples increase with decreasing 
the temperature and the averaged Josephson current agrees with  
the analytic results in Eq.~(\ref{jc1}) which are plotted in 
Fig.~\ref{result} (c),
where a parameter $L_N/\xi_D|_{T=T_c}=7.6$ is estimated from the 
numerical simulation.
 In Fig.~\ref{result}(b), the averaged Josephson current 
is almost zero for $\alpha_{L,R}=\pi/4$
as it is predicted in Eq.~(\ref{zero}).
This fact, however, does not mean the absence of the Josephson current
 in experiments for a single sample. 
The sign of the Josephson current is either
positive or negative depending on the random configuration
 and $|J|$ of a single sample increases rapidly with decreasing
the temperature as shown in Fig.~\ref{result}(b). 
These results indicate an importance of
the mesoscopic fluctuations in the Josephson
current~\cite{altshuler,beenakker2}.
In Eq.~(\ref{iso}), we assume that $\langle {\cal T}(k_y,k'_y) \rangle$ 
is independent of $k_y$ and $k'_y$. In other words, the transmission
coefficients are isotropic in the momentum space.
In order to explain the numerical results in Fig.~\ref{result}(b), 
we have to consider 
a sample-specific anisotropy in the transmission coefficients.
Here we introduce the anisotropy in a simple function,
\begin{eqnarray}
& &{\cal T}(k_y,k'_y) \rightarrow 
\overline{\langle t^e t^h \rangle} \nonumber \\
&\times& \left[ 1+f_1\; \delta_{k_y,k_1}\; \delta_{k'_y,k_1} 
+f_2 \; \delta_{k_y,k_2} \; \delta_{k'_y,-k_2} \right]
\label{imb}
\end{eqnarray}
where $f_1$ and $f_2$ are positive numerical factors much smaller than unity.
In Eq.~(\ref{imb}), we consider the situation where the two elements
${\cal T}(k_1,k_1)$ and ${\cal T}(k_2,-k_2)$ are slightly larger than the 
average. By using Eq.~(\ref{imb}), the
Josephson current in a single sample with $\alpha_{L,R}=\pi/4$ results in
\begin{eqnarray}
{\bar J}(\pi/4,\pi/4) &=& \frac{\Delta}{\Delta_0}
T\sum_{\omega_n} 
\frac{ ln}{\sinh ln}\nonumber \\
&\times&\left[ f_1 Y(\gamma_1)- f_2Y(\gamma_2)\right],
\label{jc3}
\\
Y(\gamma)&\equiv& \frac{3}{4}\frac{\Delta \cos^6 \gamma \sin^2 2\gamma}
{\left[|\omega_n|+\frac{\Delta}{2Z^2}\cos^2 \gamma |\sin 2\gamma |\right]^2}.
\label{yg}
\end{eqnarray}
The denominator of Eq.~(\ref{yg}) approaches to zero
in the limit of $\omega_n \rightarrow 0$ and $Z \gg 1$, which 
reflects the ZES at the NS interface. 
In Fig.~\ref{result} (d), we plot the analytical results in Eq.~(\ref{jc3})
for several choices of $f_1$, $f_2$, $\gamma_1$ and $\gamma_2$, where
$f_1$ and $f_2$ are of the order of $10^{-2}$. 
Some of the results show non-monotonic temperature dependence, 
and others change the sign with decreasing the temperature
as shown in both Figs.~\ref{result}(b) and (d).
The analytical results well explain the numerical results.
These results indicate the enhancement of the mesoscopic fluctuations
of the Josephson current in the limit of $T\rightarrow 0$.

\section{Discussion}

In Sec.III, we calculate ${\cal T}(k_y,k'_y)$ by solving the diffusion
equation in Eq.~(\ref{diff}).
The conclusion in Eq.~(\ref{zero}) is independent of the detail of the
approximation. In the diffusive conductors,  
$\langle {\cal T}(k_y,k'_y)\rangle$ are independent of $k_y$ and $k'_y$.
The conclusion in Eq.~(\ref{zero}) always holds, when 
$\langle {\cal T}(k_y,k'_y)\rangle$ is isotropic in the momentum space
because the integration with respect to $\gamma$ at the two NS interfaces 
can be carried out independently in Eq.~(\ref{jgene}). 
For this reason, we derive an equation for the dirty SNS junctions,
\begin{equation}
\langle J(\alpha_L,\pi/4) \rangle = \langle J(\pi/4,\alpha_R) \rangle 
=0
\end{equation}
for any $\alpha_L$ and $\alpha_R$.
We also derive relations,
\begin{eqnarray}
& &\langle J(\alpha_L,\alpha_R) \rangle = \langle J(\alpha_L,-\alpha_R) 
\rangle \nonumber \\
&=& \langle J(-\alpha_L,\alpha_R) \rangle
= \langle J(-\alpha_L,-\alpha_R) \rangle
\end{eqnarray}
because the zero energy states do not contribute to the ensemble average
of the Josephson current.  

In the $s$-wave SNS junctions, the amplitude of the
fluctuations is $\delta J_c=(e E_{c}/\hbar) C_d$, where 
$E_c=\hbar D_0/L_N^2$ is the 
Thouless energy of the normal conductor and $C_d$ is a constant 
of the order of unity~\cite{altshuler}. 
The amplitude of the fluctuations in the Josephson current 
for $\alpha_{L,R}=0$ at $T=0$ are expected to be proportional to the 
Thouless energy and depend on the sample size
since the characteristic features 
of the Josephson current are essentially the same with 
those in the $s$-wave junctions. 
On the other hand, the fluctuations
for $\alpha_{L,R}=\pi/4$ in Fig.~\ref{result}(b) are mainly due to the 
ZES at the interfaces and strongly depend on $Z$ and $T$ as shown in 
Eq.~(\ref{yg}) and Fig.~\ref{result}(d).
Though the average of the Josephson current is zero, the fluctuations
for $\alpha_{L,R}=\pi/4$ can be larger than those for $\alpha_{L,R}=0$.
Further theoretical investigations are necessary to understand the 
amplitude of the fluctuations.

The impurities in the normal metal may suppress the  
pair potentials near the NS interfaces
because the normal impurities break a Cooper pair in the anisotropic
superconductors. 
The pair potential should be determined self-consistently in such situation.
So far in SIS junctions, 
Tanaka, et. al.~\cite{tanaka2} calculated the Josephson current where
the pair potentials determined in a self-consistent way and compared
with the Josephson current obtained in a non-self-consistent manner. 
Their results show that there are no qualitative differences 
in the Josephson current between the two methods.
In their study, the pair-breaker is the insulator. 
The pair-breaker in the present paper is the impurities in the normal
metal. Though the origin of the pair-break are different in the 
two systems, the suppression of the pair potentials, therefore, the 
suppression of the Josephson current is considered to be the 
common consequence when we determine the pair potential self-consistently. 
Thus it may be possible to infer that the Josephson current in real
SNS junctions would be smaller than that of the present paper.
The self-consistent study has to be done to discuss the amplitude 
of the Josephson current quantitatively.
 However the main conclusion in Eq.~(\ref{zero}) is not sensitive to 
the profile of the pair potential 
because the $d$-wave symmetry is responsible for the disappearance of the 
averaged Josephson current. The impurities may also modify the symmetry of 
the pair potentials. If the finite values of 
averaged Josephson current is observed in experiments, such results
might be an evidence for the change of the pairing symmetry due to 
the impurities. This is because the $d$-wave component of the pair 
potentials does not contribute to the averaged current for 
$\alpha=\pi/4$.

The sign change of the pair potential on the Fermi surface, 
which is a characteristic feature of the anisotropic superconductors,
leads to the disappearance of the Josephson current. Thus
the conclusion in Eq.~(\ref{zero}) may be applied to the SNS
junctions of the superconductors with $p$-wave pairing symmetry
such as UPt$_3$ for certain orientation angles between the junction
interface normal and the node lines of the pair potentials. 
The investigation in this direction is now in progress.

\section{Conclusion}
 In conclusion, we analytically derive the general expression of the 
Josephson current in SNS junctions of the $d$-wave superconductors.
In dirty SNS junctions, we show that the ZES do not contributes to the
ensemble-averaged Josephson current because of the
symmetry of the pair potential.
In particular, when the orientation angle of the $d$-wave superconductor 
is $\alpha=\pi/4$, the ensemble-averaged Josephson current vanishes.
 The critical current of a single sample, however, 
remains finite value, which indicate
an importance of the mesoscopic fluctuations of the Josephson 
current.

\section*{Acknowledgements}
The author is indebted to N.~Tokuda, H.~Akera, Y.~Tanaka
and Y.~Takane for useful discussion. 

\appendix
\section{Transmission and Reflection Coefficients}

We derive the transmission and reflection coefficients
of the NS interface at $x=0$ as shown in Fig.~\ref{app1}. 
In what follows, we calculate the coefficients after analytic
continuation ($E\rightarrow i\omega_n$) for $\omega_n > 0$
and we assume that $\Delta \ll \mu_F$. 
In the normal metal, the wavefunction of the quasiparticle 
with $k_y=k_F\sin\gamma$ 
can be described by
\begin{eqnarray}
& &\hat{\Psi}^N_{k_y}(x,y) = \left[
\bar{\alpha} \left( \begin{array}{c} 1 \\ 0 \end{array} \right)
{\rm e}^{-{\rm i}p_+x}
+\bar{\beta} \left( \begin{array}{c} 0 \\ 1 \end{array} \right)
{\rm e}^{{\rm i}p_-x}
\right. \nonumber \\
&+& \left.
\bar{A} \left( \begin{array}{c} 1 \\ 0 \end{array} \right)
{\rm e}^{{\rm i}p_+x}
+ \bar{B} \left( \begin{array}{c} 0 \\ 1 \end{array} \right)
{\rm e}^{-{\rm i}p_-x}
\right] \chi_{k_y}(y),
\end{eqnarray}
where $\bar{\alpha}$ and $\bar{\beta}$ ($\bar{A}$ and $\bar{B}$ ) 
are the amplitudes of the 
incoming (outgoing) waves
in the electron and hole channels, respectively.
The wavefunction in the $y$ direction is $\chi_{k_y}(y)$,
$p_\pm= k_F\cos\gamma \pm {\rm i} \omega_n/v_F \cos\gamma$, and
$v_F=k_F/m$ is the Fermi velocity, respectively. 

In the same way, the wavefunction in the superconductor is
given by %
\begin{eqnarray}
 &\hat{\Psi}^S_{k_y}&(x,y) = \left[
\bar{\gamma} {\Phi}^+_L 
\left( \begin{array}{c} u^+_L \\ v^+_L \end{array} \right)
{\rm e}^{{\rm i}k_+ x}
+\bar{\delta} {\Phi}^-_L 
\left( \begin{array}{c} v^-_L \\ u^-_L \end{array} \right)
{\rm e}^{-{\rm i}k_-^\ast x}
\right. \nonumber \\
&+&  \bar{C} {\Phi}^-_L 
\left( \begin{array}{c} u^-_L \\ v^-_L \end{array} \right)
{\rm e}^{-{\rm i}k_- x}
+  \left.\bar{D} {\Phi}^+_L 
\left( \begin{array}{c} v^+_L \\ u^+_L \end{array} 
\right)
{\rm e}^{{\rm i} k_+^\ast x}
\right]  \nonumber \\
& &\times \chi_{k_y}(y),
\end{eqnarray}
where $\bar{\gamma}$ and $\bar{\delta}$ ($\bar{C}$ and $\bar{D}$ )
are the amplitudes of incoming (outgoing) waves
in the electron and hole channels, respectively. 
We define that 
\begin{eqnarray}
\Delta^\pm_j&=& \Delta \cos2(\gamma \mp \alpha_j), \\
\Omega^\pm_j&=&\sqrt{\omega_n^2+{\Delta^\pm_j}^2},\\ 
u^\pm_j &=& \sqrt{ \left(1 + \Omega^\pm_j/\omega_n\right)/2}, \\
v^\pm_j &=& \sqrt{ \left(1 - \Omega^\pm_j/\omega_n\right)/2}, \\
k_\pm &=& k_F \cos\gamma + i \Omega^\pm_j / v_F \cos\gamma,
\end{eqnarray}
where $j=L$ or $R$.
The phase and the sign of the pair potential is considered in the matrix,
\begin{equation}
{\Phi}^\pm_j = \left( \begin{array}{cc} 
{\rm e}^{{\rm i} (\varphi_j+\bar{\phi}_\gamma^\pm)/2} & 0 \\
0 & {\rm e}^{-{\rm i}(\varphi_j+\bar{\phi}_\gamma^\pm)/2}
\end{array}\right),
\end{equation}
where ${\rm e}^{i \bar{\phi}_\gamma^\pm} = sgn( \Delta^\pm)$.
The two wavefunctions satisfy the continuity condition at the left
NS interface (i.e., $x=0$ and $\alpha_j=\alpha_L$),
\begin{eqnarray}
\hat{\Psi}^N_{k_y}(0,y) &=& \hat{\Psi}^S_{k_y}(0,y), \label{conti1}\\
 \left. \frac{\partial}{\partial x} \hat{\Psi}^N_{k_y}(x,y)\right|_{x=0}
&=&  \left. \frac{\partial}{\partial x} \hat{\Psi}^S_{k_y}(x,y)\right|_{x=0}
\nonumber \\
& &+ 2mV_b \hat{\Psi}^S_{k_y}(0,y) \label{conti2}.
\end{eqnarray} 
From Eqs.~(\ref{conti1}) and (\ref{conti2}), the amplitudes of the outgoing
waves are connected with those of the incoming waves.
\begin{equation}
\left(\begin{array}{c}
\bar{A} \\  \\ \bar{B} \\ \\ \bar{C} \\ \\ \bar{D} \end{array}
\right)
= \left( 
\begin{array}{cccc}
r_{NN}^{ee} & r_{NN}^{eh} & t_{NS}^{ee} & t_{NS}^{eh} \\ \\
r_{NN}^{he} & r_{NN}^{hh} & t_{NS}^{he} & t_{NS}^{hh} \\ \\ 
t_{SN}^{ee} & t_{SN}^{eh} & r_{SS}^{ee} & r_{SS}^{eh} \\ \\
t_{SN}^{he} & t_{SN}^{hh} & r_{SS}^{he} & r_{SS}^{hh} 
\end{array}
\right)
\left(
\begin{array}{c}
\bar{\alpha} \\ \\ \bar{\beta} \\ \\ \bar{\gamma} \\ \\ \bar{\delta} 
\end{array}
\right). \label{mat1} 
\end{equation}
We explicitly show 10 coefficients which are required for calculating 
the Josephson current,
\begin{eqnarray}
t_{NS}^{ee} &=&-{\rm i}M^\ast \frac{\Delta^-_LK^+_L}{v^+_L} 
\frac{\Omega^+_L}{\omega_n}
{\rm e}^{-{\rm i}\frac{\bar{\phi}_\gamma^+}{2}}
{\rm e}^{{\rm i}\frac{\varphi_L}{2}} , \\
t_{SN}^{hh} &=&-{\rm i}M \;\frac{\Delta^-_LK^+_L}{v^+_L} 
\phantom{\frac{\Omega^+_L}{\omega}}
 {\rm e}^{-{\rm i}\frac{\bar{\phi}_\gamma^+}{2}}
{\rm e}^{{\rm i}\frac{\varphi_L}{2}}, \\
t_{NS}^{he} &=&-N \; \frac{\Delta^+_LK^-_L}{u^+_L} 
\frac{\Omega^+_L}{\omega_n}
{\rm e}^{{\rm i}\frac{\bar{\phi}_\gamma^+}{2}}
{\rm e}^{-{\rm i}\frac{\varphi_L}{2}}, \\
t_{SN}^{he} &=&\phantom{-}N \frac{\Delta^+_LK^-_L}{u^+_L} 
\phantom{\frac{\Omega^+}{\omega}}
 {\rm e}^{{\rm i}\frac{\bar{\phi}_\gamma^+}{2}}
 {\rm e}^{-{\rm i}\frac{\varphi_L}{2}}, \\
t_{NS}^{hh} &=&-{\rm i}M \; \frac{\Delta^+_LK^-_L}{v^-_L} 
\frac{\Omega^-_L}{\omega_n}
{\rm e}^{{\rm i}\frac{\bar{\phi}_\gamma^-}{2}}
{\rm e}^{-{\rm i}\frac{\varphi_L}{2}} ,\\
t_{SN}^{ee} &=&-{\rm i}M^\ast \frac{\Delta^+_LK^-_L}{v^-_L}
\phantom{\frac{\Omega^+}{\omega}}
{\rm e}^{{\rm i}\frac{\bar{\phi}_\gamma^-}{2}}
{\rm e}^{-{\rm i}\frac{\varphi_L}{2}} , \\
t_{NS}^{eh} &=&\phantom{-}N \frac{\Delta^-_LK^+_L}{u^-_L}
\frac{\Omega^-_L}{\omega_n}
 {\rm e}^{-{\rm i}\frac{\bar{\phi}_\gamma^-}{2}}
{\rm e}^{{\rm i}\frac{\varphi_L}{2}}, \\
t_{SN}^{eh} &=&-N \; \frac{\Delta^-_LK^+_L}{u^-_L} 
\phantom{\frac{\Omega^+}{\omega}}
{\rm e}^{-{\rm i}\frac{\bar{\phi}_\gamma^-}{2}}
{\rm e}^{{\rm i}\frac{\varphi_L}{2}}, \\
r_{NN}^{he}&=& -{\rm i} Q(-k_y,\alpha_L)
{\rm e}^{-{\rm i}\varphi_L}, \\
r_{NN}^{eh}&=& -{\rm i}Q(k_y,\alpha_L){\rm e}^{{\rm i}\varphi_L}, \\
M&\equiv& \frac{\cos\gamma(\cos\gamma + {\rm i}Z)}{ \Xi_L}, \\
N&\equiv& \frac{ Z \cos\gamma}{ \Xi_L}, \\
{\Xi}_j&\equiv&
Z^2(\Delta^+_j\Delta^-_j + K^+_jK^-_j) +\cos^2\gamma \Delta^+_j\Delta^-_j,\\
Z&\equiv& \frac{mV_b}{k_F},\\
K^\pm_j&\equiv& \Omega^\pm_j-|\omega_n|, \\
Q&(&k_y,\alpha_j) \equiv \frac{\cos^2\gamma \Delta^-_jK^+_j}{\Xi_j}, \\
Q&(&-k_y,\alpha_j) \equiv \frac{\cos^2\gamma \Delta^+_jK^-_j}{\Xi_j}.
\end{eqnarray}
These coefficients are a function of $k_y$, $\alpha_L$ and $\varphi_L$.
On the derivation, we use the approximation 
$k_\pm = p_\pm = k_F \cos\gamma$, where we assume that $\omega_n \ll \mu_F$.

The reflection coefficients at the right NS interface are obtained in the same 
way. The two Andreev coefficients are given by
\begin{eqnarray}
r_{NN}^{he}(k_y,\alpha_R,\varphi_R) 
&=& -{\rm i}Q(k_y,\alpha_R) {\rm e}^{-{\rm i}\varphi_R}, \\
r_{NN}^{eh}(k_y,\alpha_R,\varphi_R)
 &=& -{\rm i}Q(-k_y,\alpha_R){\rm e}^{{\rm i}\varphi_R}. 
\end{eqnarray}

\section{Solution of the Diffusion Equation}
In this paper, we consider the low temperature regime 
$T<T_c\sim 10^{-3}-10^{-4} \mu_F$ and the diffusive
transport regime $1/\mu_F\tau_0 \sim 10^{-1}-10^{-2}$.
Thus a relation $2\pi T \tau_0 < 1 $ is satisfied.
In these regime, we solve the diffusion equation 
\begin{equation}
\tau_0\left(2|\omega_n|-D_0 \nabla^2_{\bf r} \right)X({\bf r},{\bf r}')
\approx  2\pi N_0\tau_0 \delta({\bf r}-{\bf r}'), \label{diffa}
\end{equation}
under the appropriate boundary condition~\cite{kresin,lee}.
The right hand side of Eq.~(\ref{diffa}) corresponds to
the first term in Fig.~\ref{diffusion} which can be 
calculated to be
\begin{equation}
\frac{N_0}{v_F r} {\rm e}^{-r/\ell},
\end{equation} 
where $\ell=v_F\tau_0$ is the mean free path and $r=|{\bf r}-{\bf r}'|$,
respectively. In Eq.~(\ref{diffa}), we replace this function by 
the $\delta$-function in two-dimension because the smallest 
length scale in the diffusion
equation is $\ell$. 
The function $X({\bf r},{\bf r}')$ is the Green function 
of the Schr\"{o}dinger-type equation
\begin{equation}
D_0\nabla^2 f({\bf r})=\lambda f({\bf r}).\label{sch}
\end{equation}
The normal conductor is separated by the high potential barrier
at the two NS interfaces and there is no current flow in the $y$ direction.
In such situation, we solve Eq.~(\ref{sch}) under the boundary
conditions,
\begin{eqnarray}
\left.\frac{ \partial f({\bf r})}{\partial x}\right|_{x=0,L_N}&=&0, 
\label{bcx}\\
\left.\frac{ \partial f({\bf r})}{\partial y}\right|_{y=0,W}&=&0. 
\end{eqnarray}
The function $X({\bf r},{\bf r}')$ is represented by
\begin{eqnarray}
& &X({\bf r},{\bf r}')= 2\pi N_0
\sum_{m=0}^\infty \sum_{m'=0}^\infty A_m^2 B_{m'}^2\nonumber \\
&\times&
\frac{\cos\left( \frac{m\pi x}{L_N}\right) 
\cos\left(\frac{m\pi x'}{L_N}\right)
\cos\left( \frac{m'\pi y}{W}\right) 
\cos\left(\frac{m'\pi y'}{W}\right)}
{2|\omega_n|+D_0[(m\pi/L_N)^2+(m'\pi/W)^2]}, \\
& &A_m = \left\{\begin{array}{cc}
\sqrt{\frac{1}{L_N}},& \hbox{for} \: m=0 \\
\sqrt{\frac{2}{L_N}},& \hbox{for} \: m\neq 0,
\end{array}\right. \\
& &B_m = \left\{\begin{array}{cc}
\sqrt{\frac{1}{W}},& \hbox{for} \: m=0 \\
\sqrt{\frac{2}{W}},& \hbox{for} \: m\neq 0.
\end{array}\right.
\end{eqnarray}
Next we calculate the function
\begin{eqnarray}
& &\left. \int_{0}^W dy \int_{0}^W dy' 
X({\bf r},{\bf r}')\right|_{x=L_N, x'=0}\nonumber \\
&=&2\pi N_0 \tau_0 \frac{W}{L_N}  
\frac{1}{2|\omega_n|\tau_0}\frac{ln}{\sinh ln},
\end{eqnarray} 
where $ln = \sqrt{2n+1} L_N/\xi_D$ and $\xi_D=\sqrt{D_0/2\pi T}$ is the
coherence length.
Finally the averaged transmission coefficients are
\begin{equation}
\overline{\langle t^e t^h \rangle} = \left(\frac{1}{N_c}\right)^{2} 
g_N \frac{ln}{\sinh ln} \frac{1}{2|\omega_n|\tau_0},\label{sing}
\end{equation} 
where $g_N=2\pi N_0 D_0 W/L_N$ is the dimensionless conductance
of the normal metal for each spin degree of freedom.
 In Eq.~(\ref{sing}), $\overline{\langle t^e t^h \rangle}$ 
seems to be singular in the limit of $\omega_n\rightarrow 0$.
We note that the singularity is stemming from the boundary
condition in Eq.~(\ref{bcx}). When there is no potential 
barrier at the NS interface, we apply the boundary condition
$f({\bf r})|_{x=0,L_N}=0$ instead of Eq.~(\ref{bcx}) and obtain
\begin{eqnarray}
\overline{\langle t^e t^h \rangle}&=& \left(\frac{1}{N_c}\right)^{2} 
g_N \frac{ln}{\sinh ln} \nonumber \\ 
&\times& \frac{1}{2|\omega_n|\tau_0} 
\frac{\cosh( 2\sqrt{2|\omega_n|\tau_0})-1}{2},
\end{eqnarray}
where the integration with respect to $y$ and $y'$ are carried out
at $x=L_N+\ell/\sqrt{2}$ and $x'=-\ell/\sqrt{2}$.
Since the propagation of a quasiparticle in the normal conductor
is not sensitive to the boundary condition at the NS interface and 
$N_c^2\overline{\langle t^e t^h \rangle}$ is close to $g_N$
in the limit of $\omega_n\rightarrow 0$, we regularize the
Eq.~(\ref{sing}) by introducing the cut-off,
\begin{eqnarray}
\overline{\langle t^e t^h \rangle} &=& \left(\frac{1}{N_c}\right)^{2} 
g_N \frac{ln}{\sinh ln} F(2|\omega_n|\tau_0),
\\
F(x) &=& \Theta(1-x) + \Theta(x-1)\frac{1}{x},
\end{eqnarray} 
where $\Theta(x)$ is the step function.
Since $\overline{\langle t^e t^h \rangle}$ decays as 
$\exp(-\sqrt{2n+1}L_N/\xi_D)$, the contributions
from small $n$ are dominant in the summation with respect to $\omega_n$
in Eq.~(\ref{jsns}). Thus in most cases, $F(2|\omega_n|\tau_0)$ is unity.

* e-mail: asano@eng.hokudai.ac.jp

\begin{figure}
\caption{
The SNS junction of the $d$-wave superconductor is illustrated. 
The orientation angle in the left (right) superconductor is $\alpha_L$ 
($\alpha_R$).
}
\label{system}
\end{figure}
\begin{figure}
\caption{
The reflection coefficients in (a) contribute to the 
Josephson current proportional to $\sin(\varphi_L-\varphi_R)$.
The Josephson current calculated from the four 
reflection coefficients in (a) is summarized in a reflection
process in (b).
}
\label{process}
\end{figure}
\begin{figure}
\caption{
The propagation process in the diffusive normal conductor is 
diagrammatically illustrated. The cross represents the impurity
scattering.
}
\label{diffusion}
\end{figure}
\begin{figure}
\caption{ The Josephson current in dirty SNS junctions 
is shown as a function of the temperatures.
The numerical results for $\alpha_{L,R}=0$ and $\pi/4$ 
are plotted in (a) and (b), 
respectively. The lines are the numerical results of 
several samples with different random configurations and 
the open circles are the average over
a number of samples. 
For comparison, we show the analytical results for $\alpha_{L,R}=0$ in (c)
and for $\pi/4$ in (d), respectively. 
}
\label{result}
\end{figure}
\begin{figure}
\caption{The transmission and the reflection coefficients at the left NS
interface. There are four incoming and outgoing channels.
}
\label{app1}
\end{figure}

\end{multicols}

\end{document}